\documentclass[aps,floatfix,tightenlines,showkeys,showpacs,superscriptaddress,twocolumn]{revtex4}

\usepackage{epsfig}
\usepackage{graphicx}
\usepackage{amssymb}
\newcommand{\nn}{\nonumber\\}

\newcommand{\ben}{\begin{displaymath}}
\newcommand{\een}{\end{displaymath}}
\newcommand{\be}{\begin{equation}}
\newcommand{\ee}{\end{equation}}
\newcommand{\bea}{\begin{eqnarray}}
\newcommand{\eea}{\end{eqnarray}}

\newcommand{\eq}[1]{Eq.~(\ref{#1})}

\newcommand{\fig}[1]{Fig.~\ref{#1}}
     
 \newcommand{\bfq}{{\bf q}}
 \newcommand{\bfb}{{\bf b}}

\begin{document}

\title{\bf  \hskip12cm NT@UW-10-15\\\vskip.3cm{
Realistic  Transverse Images of the Proton Charge and Magnetic Densities }}

\author{Siddharth Venkat}
\affiliation{Virginia Polytechnic Institute and State University, Blacksburg, VA 24061-0002}
\affiliation{Department of Physics,University of Washington, Seattle, Washington 98195-1560}
                                                                           
\author{John Arrington}
\affiliation{Physics Division, Argonne National Laboratory, Argonne, Illinois 60439}

\author{Gerald A. Miller} \email[Corresponding author: ]{miller@phys.washington.edu}
\affiliation{Department of Physics,University of Washington, Seattle, Washington 98195-1560}

\author{Xiaohui Zhan}
\affiliation{Physics Division, Argonne National Laboratory, Argonne, Illinois 60439}

\date{\today}

\begin{abstract}

We develop a technique, denoted as the finite radius approximation (FRA), that
uses a two-dimensional version of the Shannon-Nyquist sampling theorem to
determine transverse densities and their uncertainties from experimental
quantities.  Uncertainties arising from experimental uncertainties on the form
factors and lack of measured data at high $Q^2$ are treated. A key feature of
the FRA is that a form factor measured at a given value of $Q^2$ is related to
a definite region in coordinate space.  
An exact relation between the FRA and
the use of a Bessel series is derived. The proton Dirac form factor is well enough known such that
 the transverse charge
density is very accurately known except for transverse separations $b$ less than
about 0.1~fm. The Pauli form factor is  well known to $Q^2$ of about 10 GeV$^2$, and this allows
a reasonable, but improvable, determination of the anomalous magnetic moment density.
\end{abstract}

\maketitle
 
\section{Introduction}\label{intro}

A truly impressive level of experimental technique, 
effort and ingenuity has been applied
to  measuring  the electromagnetic form factors
of the proton, neutron (nucleon) and pion
\cite{Gao:2003ag,HydeWright:2004gh,Perdrisat:2006hj,Arrington:2006zm,Horn:2007ug,Blok:2008jy}.
These quantities are  probability amplitudes
that a given hadron 
can absorb a specific  amount of momentum and  remain in the ground
state, 
 and therefore  should supply  information about 
charge and magnetization spatial densities.

The  text-book  interpretation of these form factors is  that their
 Fourier transforms are measurements of the charge and magnetization 
densities.
This interpretation is deeply buried in the 
thinking of nuclear  physicists 
and continues to guide intuition, as it has since the days
of the Nobel prize-winning work of Hofstadter\cite{Hofstadter:1956qs}.
Nevertheless, the relativistic motion of the constituents of the
system causes the text-book interpretation 
to be incorrect\cite{Miller:2009sg}. The difficulty  is that in
electron-proton scattering the initial and final nucleon states have different
momenta and therefore different wave functions. In general, these different
states are related by a boost operator that depends on the full complexity of
QCD. The use of transverse densities  \cite{Miller:2007uy,Miller:2010nz} avoids this difficulty by  working in the infinite
momentum frame and taking the spacelike momentum transfer to be in the
direction transverse to that of the infinite momentum. In this case,  the different momenta of the initial and final
nucleon states are  accommodated by using 
two-dimensional Fourier transforms.
The transverse charge and magnetization densities  are constructed from
density operators that are the absolute square of quark-field operators, so they are correctly defined as densities.

This paper is concerned with extracting the spatial information by developing
and using a theoretical technique that is model-independent and also provides
a practical way of dealing with both experimental uncertainties and the
lack of information on unmeasured regions, with minimal assumptions.
In the subsequent text we plan to show how to construct bands
of transverse densities that are consistent with available experimental
knowledge and also take into account the possible effects of data taken at
momentum transfer $Q^2$ higher than available in the present data set. This
allows one to consider the possible impact of future experiments.

But there also is a more general context, with the high current interest in
mapping the three-dimensional structure of the nucleon~\cite{Burkert:2008rj}. Therefore
we also aim to provide a technique that can be easily extended determining the
spatial aspects of  other quantities related to transverse momentum
distributions and generalized parton distributions.

Next we present an overview of the remainder of this paper. 
Sect.~\ref{theory} concerns the following situation. Suppose a form factor
$F(Q^2)$ and transverse density $\rho(b)$ are related by a two-dimensional
Fourier transform, and that $\rho(b)$ is localized, $\rho(b)=0$ for $b$
greater than some finite distance. The function $\rho(b)$  is band limited and
can be written as a discrete Fourier series involving  $F(Q^2)$. This result, 
known as the Nyquist-Shannon \cite{SN} sampling theorem, enables us to
associate the density at a given range of values of $b$ with a discrete value
of the momentum transfer, see  \eq{cfra} below (which we denote as the finite
radius approximation FRA). The equivalence  between  the FRA and the Bessel
series  expansion  technique is also established. A general version of the
FRA, applicable to other observable quantities, is also presented.
 
 Sect.~\ref{examples} is concerned with exploring the the validity and utility
(which depends on the number of terms needed in the discrete Fourier series)
of the FRA using examples in which the form factor is given by a monopole (M)
or dipole (D) form. Sect.~\ref{results} is concerned with the reality that the
proton electromagnetic form factors are not known as analytic functions.
Instead, form factors $G_{E,M},\;F_{1,2}$ (with uncertainties)  measured at
discrete values of $Q^2$ up to a finite maximum value $Q^2_{max}$ are known. 
This means that $\rho$  is known only within some uncertainties, and  a
technique to determine the uncertainties in $\rho$ must be developed. This is
accomplished by  using the values of $F_i\pm dF_i$ in the FRA.  Estimates of
the effects of incompleteness, arising from contributions in the unmeasured
region, $Q^2>Q^2_{max}$, are also provided.  The paper is concluded with a
brief summary.

\section{General Considerations}\label{theory}

Intuitively, we expect particles to be localized. That is, we expect densities
associated with the particle to be well approximated by functions that are
zero outside some maximum radius. This assumption, called the finite radius
approximation (FRA),  greatly simplifies the relationship between form
factors and their associated densities.

Let $\rho(b)$ be a two-dimensional  transverse density function (we later take this
to be charge or magnetization density) and let $F(Q^2)$ be the associated form
factor. The transverse density is given by  \cite{Soper:1976jc,Miller:2007uy}
\bea
& \rho(b) & = {1\over( 2\pi^2)}\int d^2q e^{-i\bfq\cdot\bfb}\;F(Q^2=\bfq^2) \cr
&& = {1\over 2\pi}\int QdQ J_0(Q b)F(Q^2),\label{rhodef}
\eea
with the azimuthal symmetry of $\rho$  obtained from the Lorentz invariant
form of $F$ in the space-like region with $q^+=0$. 
If one knows $F(Q^2)$ exactly for all values of $Q^2$, the transverse density is known immediately. However, one only knows $F(Q^2)$ within experimental uncertainties for a finite range of $Q^2$. This means that $\rho$  is known only within some uncertainties, and  it is necessary to develop a technique to determine the uncertainties in $\rho$.

We proceed by assuming  that
$\rho(b)\approx0$ for $b \geq R$, where $R$ is a finite distance.
Since the functions $\rho,F $ are  Fourier transforms,  $F$ is band-limited.
We proceed in the spirit of the Nyquist-Shannon sampling theorem. The function $\rho$
can be expanded as
\be 
\rho(b)=\sum_{n=1}^{\infty}c_n J_0(X_n\frac{b}{R}), \label{exp}
\ee
where $X_n$ is the $n$-th zero of $J_0$, and $c_n$ is given approximately  by
the formula
\bea
c_n\approx \widetilde{c}_n 
&=&\frac{1}{2\pi}\frac{2}{R^2 J_1(X_n)^2} F(Q_n^2), \label{cfra}
\eea
with
\bea Q_n  \equiv \frac{X_n}{R}.\label{qn}\eea 
The above equation \eq{cfra}, which is the two-dimensional version of \cite{SN},  is
the central formal result of this paper. 
Using this  in  \eq{exp} yields  the following expression for $\rho(b)$:
\be
\rho(b)=\frac{1}{\pi R^2}\sum_{n=1}^{\infty}J_1(X_n)^{-2}F(Q_n^2)J_0(X_n\frac{b}{R}),\label{efra}
\ee
The result \eq{efra} is the central phenomenological result because it tells us
that measuring a  form factor at $ Q_n^2$ provides information about the
density mainly at values of $b< R/X_n$. This is because Bessel functions are
of the order of unity only for values of arguments less than that of its first
zero.

\subsection{Equivalence with the Bessel Series}\label{equiv}
Replacing $c_n$ by $\widetilde{c}_n$  would be exact if  the assumption 
 $\rho(b\ge R)=0$  is
exactly true. This  condition is clearly approximately true, so we expect a near
equality between $c_n$ and $\widetilde{c}_n$. In fact, it turns out that the
approximation is amazingly accurate as we now demonstrate. Numerical examples are provided in subsequent sections. The exact values of $c_n$ are obtained
from the orthogonality of the cylindrical Bessel functions as
\be
c_n=\frac{2}{R^2 J_1(X_n)^2}\int_0^R
b\rho(b)J_0(X_n\frac{b}{R})\,db.\label{cexa}\ee The use of  this in
\eq{rhodef} followed by integration over $b$ can be done using a standard
identity to yield
\be
c_n=\frac{X_n}{\pi  R^2 J_1(X_n)}\int_0^{\infty}\frac{q F(q^2) J_0(q R)}{(\frac{X_n}{R})^2-q^2}\,dq .\label{cc}
\ee

We may use a dispersion relation for the form factor \cite{Strikman:2010pu}
 to establish the
connection between $\widetilde{c}_n$ and $c_n$. First recall that, for $Q^2>0$,
\bea F(Q^2)={1\over \pi}\int_{4m_\pi^2}^\infty \;dt {ImF(-t)\over t+Q^2 },\label{disp}\eea
and that using this expression in \eq{rhodef} yields
\bea
\rho(b)={1\over 2\pi^2}\int_{4m_\pi^2}^\infty\;dt \;K_0(\sqrt{t}b)ImF(-t).\label{rhodsp}\eea
Proceed by using the above in \eq{cexa} and then integrate over $b$ using:
\bea
\int_0^R bdbK_0(\sqrt{t}b)J_0(b{X_n\over R}) = {1\over {X_n^2\over R^2}+t} \times \cr
\left[1+X_nJ_1(X_n)K_0(\sqrt{t} R)\right].
\eea
Then,
\bea
& c_n & =\frac{1}{R^2J_1(X_n)^2\pi^2}\int_{4m_\pi^2}^\infty \;dt {ImF(-t)\over t+{X_n^2\over R^2} } \times \cr
&&\left[1+X_nJ_1(X_n)K_0(\sqrt{t} R)\right] .
\eea
Using only the first term within the brackets along with   \eq{disp}  allows
one to identify the integral over $t$ as $\pi F(Q_n^2)$. Thus  (using
\eq{cfra}) one arrives at the result that $c_n=\widetilde{c}_n$ plus a
correction term,  suppressed by  a modified Bessel function evaluated at  a
large argument. For example, a significant contribution to the $Im\;F$ comes
from the region $t\sim m_\rho^2=0.5$ GeV$^2$, and using $R=3.3$  fm (see
Sec.~\ref{examples}), then $K_0(m_\rho R)= 10^{-6}$. The net result is that
\bea && c_n 
=\widetilde{c}_n+\delta_n\cr
&&\delta_n\equiv\frac{X_nR^2 J_1(X_n)}{J_1^2(X_n)\pi^2}\int_{4m_\pi^2}^\infty \;dt {ImF(-t)\over t+{X_n^2\over R^2} }K_0(\sqrt{t} R).\cr&&\label{delta}
\eea
A reasonable estimate   is that 
\bea
\frac{c_n-\widetilde{c}_n}{\widetilde c_n}\sim  X_nJ_1(X_n) 10^{-6}.\label{frac}
\eea
The condition that $\delta_n$ be small is that $R$ be chosen to be large
enough. We ensure that this condition is well-satisfied for all of our
examples and applications.

\subsection{Preliminary Evaluations}
It is worthwhile to perform some preliminary analysis of the  expression \eq{efra}.
For $x \gg 1,\;J_0(x)$ is well approximated   \cite{Jackson} by 
\bea
J_0(x)\approx \sqrt{\frac{2}{\pi x}}\cos(x-\frac{\pi}{4}) ~,
\eea
so that the $n$'th zero of $J_0$, $X_n$, is given approximately by
\bea X_n \approx (n+\frac{3}{4})\pi ~,
\eea
and
\bea&&
J_1(X_n)=-J_0'(X_n)\cr &&\approx         (-1)^n2^{1/2}\pi^{-1}((n+\frac{3}{4}))^{-1/2}.\label{j1Xn}
\eea
It follows that for large $n$, the terms in the series \eq{efra} for $\rho(b)$
are of the form:
\[
\frac{\pi}{2 R^2}(n+\frac{3}{4})F(Q_n^2)J_0(X_n\frac{b}{R})\sim n\;F(({n\pi\over R})^2)
\]
at $b=0$.
So for the series to converge everywhere, namely at b=0, we need $F$ to fall
faster than $Q^{-2}$ for large $Q$.  The oscillations of the cylindrical Bessel functions hastens the convergence for non-zero values of $b$.

Given this convergence, the function $\rho(b)$ can be approximated by using a finite number of terms
in the series \eq{efra}. Because $Q_n^2=(X_n /{R})^2$ serves as the $Q^2$ in the
argument of $F$, cutting off the series at $N$ terms is equivalent to taking
$F(Q^2)=0$ for $Q^2>(X_N/R)^2$.

If the assumption that $\rho(b)=0$ for $b \geq R$ holds for a given value of
$R$, then it also holds for larger values of $R$. We can see from \eq{efra}
that increasing $R$ increases the frequency with which $F(Q^2)$ is sampled and
therefore decreases the range that is sampled. As a consequence, an increase
in $R$ demands an increase in the number of terms in the approximation for
$\rho$.

A quick result following from the fact that $\rho$ is the Fourier transform of
$F$ is that the mean-square-radius $\langle b^2 \rangle$ is given by
\bea
\langle b^2 \rangle\equiv \int d^2b\;b^2\rho(b)=-4 \left.\frac{d\log F}{dQ^2}\right|_{Q^2=0}. \label{msr}
\eea
In this paper, we choose $R \approx 5\sqrt{| \langle b^2 \rangle |}$ in
determining the number of terms in our expansion. Numerical studies of the
form factors considered in preparing this paper have shown that this value of
$R$ is sufficiently large so that perturbations to this value lead to the
same density functions and that  $R^2\rho(R)$ is always small enough so that the difference between $c_n$ and
$\widetilde{c}_n$ is minute.

\subsection{Other transverse densities}\label{odt}
We believe that the techniques used in this paper can be exploited to image
other quantities that depend on transverse position. Suppose there is a
transverse quantity $\rho^{(\lambda)}(b)$ that is a two-dimensional Fourier
transform of an experimental observable $F^{(\lambda)}(Q^2)$ such that \bea
\rho^{(\lambda)}(b)={1\over2\pi}\int QdQJ_\lambda(Qb)F^{(\lambda)}(Q^2).\eea
An example, discussed in detail in Sect.~\ref{subsrhom}, is the magnetization
density $\rho_m$ of the anomalous magnetic moment. We expect that the index
$(\lambda)$ is associated with a given number of units of orbital angular
momentum. Extracting $\rho^{(\lambda)}(b)$ is facilitated by using the
expansion \bea \rho^{(\lambda)}(b)=\sum_{n=1}^\infty
c_{n\lambda}J_\lambda(X_{\lambda,n}{b\over R}),\label{gen}\eea where
$X_{\lambda,n}$ is the $n$'th zero of the Bessel function of order $\lambda$.
Then the sampling theorem leads immediately to the result. \bea&&
c_{n,\lambda}\approx \widetilde{c}_{n,\lambda}={2\over R^2
J_{\lambda+1}(X_{\lambda,n})^2}F^{(\lambda)}(Q_{\lambda,n}^2),\label{gencn}\\
&& Q_{\lambda,n}={X_{\lambda,n}\over R}\nonumber\eea The difference between $
c_{n,\lambda}$ and $ \widetilde{c}_{n,\lambda}$ can be shown to be very small
by using the arguments of Sect.~\ref{equiv}. The result \eq{gencn} can be used
to relate accessible kinematic ranges with transverse regions.

\section{Examples}\label{examples}

To demonstrate our method and explore its limitations, we now analyze two
models of the form factor. For the first model, let the form factor be given
by the monopole form
\bea
F_{M}(Q^2)=\frac{1}{1+\frac{Q^2}{\Omega^2}}\label{mpole}
\eea
where $\Omega=0.77$ GeV.  This form factor is taken as a caricature of the pion
electromagnetic form factor. Then the associated charge density is obtained
from \eq{rhodef}:
\bea \rho_{M}(b)={1\over 2\pi}
\Omega^2K_0(\Omega b).
\eea
This function diverges as $\log (1/b)$ for small values of $b$ and so provides
a severe test of the method. With the stated value of $\Omega$ we find
$\langle b^2\rangle_{M}=4/\Omega^2= 0.26\;$fm$^2$, and thus take 
$R=5\sqrt{| \langle b^2 \rangle |}=2.56$ fm. We then find
 the fractional difference between $c_n$ and $\widetilde{c}_n$ of \eq{delta}
is less than 5$\times 10^{-4}$ for small values of $n$, and the magnitude decreases rapidly as $n$ increases.

\begin{figure}[htb]
\epsfig{file=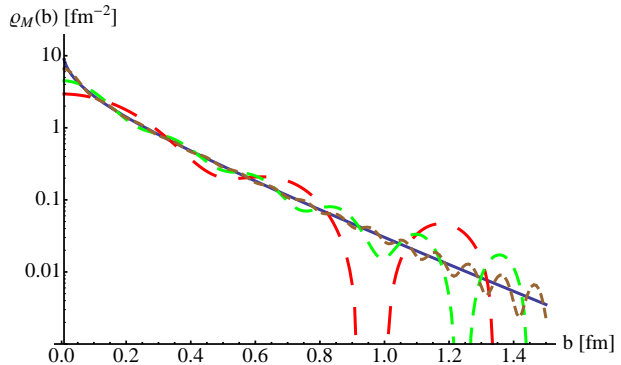, width=.48\textwidth}
\caption{(Color Online) Plot of $\rho_{M}$ (blue,solid), 10 term
approximation (red, long dash) and 20 term approximation (green, medium dash)
and 50 term (brown, short dash).}
\label{toyrhographs}\end{figure}

We  compare to $\rho_{M}$ to its  approximation as an expansion in $N$ terms,
with $N=10,20,50$ in \fig{toyrhographs}. We  see that  our approximations 
differ  from the exact result, but the difference decreases with increasing
value of $N$. The 50 term approximation works reasonably well for all value of
$b$  for which the density differs appreciably from 0. Unfortunately  the
10,20 and 50 term approximations would require  measurements at $Q^2=6,23 $
and 144 GeV$^2$. Only the first value seems presently achievable.

\begin{figure}[htb]
\epsfig{file=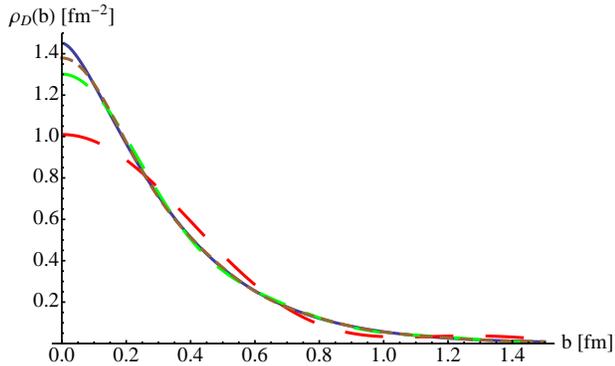, width=.48\textwidth}
\caption{(Color online) Plot of $\rho_D$ (solid), 5 term approximation (red, long dash), 10
term approximation (green, medium dash) and 15 term approximation (brown, short
dash).}
\label{rhodgraphs}\end{figure}

We now examine the dipole form factor given by
\be
F_{D}(Q^2)=\frac{1}{(1+\frac{Q^2}{\Lambda^2})^2} \label{Gd}
\ee
where $\Lambda^2=0.71$ GeV$^2$. This value is suggested by its historically
close relationship with the proton electromagnetic form factors. The dipole
transverse charge density is obtained by from \eq{rhodef} to be
\bea
\rho_{D}(b)=\frac{1}{4 \pi} b\Lambda^3K_1(b\Lambda).\label{rhoD}
\eea 
This form factor falls more rapidly with increasing $Q^2$ than does $F_M$, and
also corresponds to the larger physical extent of the proton as compared to
the pion. Furthermore, $\rho_D$ is not singular at the origin  ($\sim1-0.058
(b\Lambda)^2$). Thus there are several reasons to  expect to find better
convergence properties, and therefore a more accurate representation of the
transverse density for the proton.  With this value of $\Lambda$, $\langle
b^2\rangle=8/\Lambda^2=0.439 \;{\rm fm}^2,\;{\rm and}\;R=3.31$ fm. Once again
the fractional difference of \eq{delta} is truly tiny for all values of $n$: the fractional  differences are less than about $10^{-5}$ for all values of $n$ that correspond to non-zero $c_n$.
We plot $\rho_D$ and its approximations in Fig.~\ref{rhodgraphs}.

We can see how the approximations converge  to the exact $\rho_{D}$. Even
the 10 term approximation is reasonably good and the 15 term approximation is
extremely accurate except for $b<0.1$~fm.

Another way of looking at convergence properties is to examine properties of
the transverse density. We display upper limit dependence of $\langle b^2
\rangle_{M,D}$ for both the monopole and dipole form factors. We compute these matrix elements for a
range of values of $b$ from 0 to 1.5 fm. 
This covers the region up to where $\rho$ is about
0.1\%  of its central value. The results are shown in Table~\ref{convergance}.  Despite the relatively poor convergence obtained for the monopole form factor 
(Fig.~\ref{toyrhographs}), reasonable convergence for the expectation value is
obtained.
\begin{table}[htb]
\begin{center}
\caption{Upper limit, $N$, dependence of  $\langle b^2 \rangle_{M,D}$ computed for values of $b$ from 0 to 1.5 fm.}
\label{convergance}
\begin{tabular}{|l|c|l|}
\hline
$N$	& 		 $\langle b^2 \rangle_M\;(fm)^2$& $\langle b^2 \rangle_D\;(fm)^2$\\\hline
5	& 	 0.259&\quad 0.313\\
10	& 	0.362 &\quad0.320\\
15	& 	0.368 &  \quad0.319\\
$\infty$ &0.367   & \quad        0.319  \\
 \hline
\end{tabular}
\end{center}
\end{table}
However, the convergence is much better for the dipole form factors. The 5,10  and 15 term
approximations correspond to values of $Q^2=0.9,4$ and 9  GeV$^2$. These
values and even higher have already been achieved experimentally. Thus we
reasonably expect that the proton transverse density is now known. Indeed,
this has already been suggested \cite{Miller:2007uy}. However, now we can
answer the question: ``How well is the proton transverse charge density
known?".

\section{Extraction of Proton Form Factors and Uncertainties}\label{fitting}

The transverse densities we seek are given in terms of the Dirac $F_1$ and Pauli $F_2$ form factors, which are expressed in terms of 
 the Sachs electromagnetic form factors
$G_E$ and $G_M$ as
\be
F_1(Q^2)=\frac{G_E+\tau G_M}{1+\tau} ~,
F_2(Q^2)=\frac{G_M-G_E}{1+\tau}\label{F1F2} ~,
\ee
where $\tau=\frac{Q^2}{4 M_p^2}$.

Elastic electron-proton scattering has been measured up to $Q^2$ of about
30~GeV$^2$, with the separation of both $G_E$ and $G_M$ extracted using a
variety of techniques up to 10~GeV$^2$.  There are two sources of uncertainty
in the extraction of the transverse densities.  Experimental uncertainties
from the measurements of $G_E$ and $G_M$ yield uncertainty in the extracted
densities, and incompleteness error arise from the lack of form factor
measurements at very high $Q^2$ (above 30~GeV$^2$).  In this section, we
perform extractions of the transverse density and evaluate the
the effects that two kinds of uncertainties on the densities.

The form factors $G_E$ and $G_M$ have been extracted from a global analysis of
the world's cross section and polarization data, including corrections for
two-photon exchange corrections from Ref.~\cite{Blunden:2005ew}. The analysis
is largely identical to that that of Ref.~\cite{Arrington:2007ux},
although additional high $Q^2$ form factor results~\cite{Puckett:2010ac} have
been included.  In addition, the slopes of $G_E$ and $G_M$ at $Q^2$=0 were
constrained in the global fit based on a dedicated analysis of the low $Q^2$
data.  In the global fit, the large body of high $Q^2$ data, especially for
$G_M$, can constrain the fit well enough that the low $Q^2$ behavior is not
primarily constrained by the low $Q^2$ data. Constraining the slope
based on an analysis of only the low $Q^2$ data keeps the global fit from 
doing a poor job at low $Q^2$ simply to make a slight improvement in the
high $Q^2$ data.  In writing $G_E(Q^2)=1-Q^2R_E^2/6$, the value of
$R_E$ was constrained to be 0.878~fm and $R_M$ was constrained to be 0.860~fm.
This is important in the extraction of the large scale structure of the
density.  The fit is of the following form:
\bea
G_M(Q^2)=\mu_p\frac{1+p_6\tau+p_{10}\tau^2+p_{14}\tau^3}{1+p_2\tau+p_4\tau^2+p_8\tau^3+p_{12}\tau^4+p_{16}\tau^5} \cr
G_E(Q^2)=\frac{1+q_6\tau+q_{10}\tau^2+q_{14}\tau^3}{1+q_2\tau+q_4\tau^2+q_8\tau^3+q_{12}\tau^4+q_{16}\tau^5}\label{GeGm} \cr
\eea
where the fitting constants $p_2,..p_{16},q_2,...,q_{16}$ are given in
Table~\ref{gegmfit} and we use $\mu_p=2.792782$.

\begin{table}[htb]
\begin{center}
\caption{Fit parameters for $G_M\;(p_i)$, $G_E\;(q_i)$}\label{gegmfit}
\begin{tabular}{|l|l|l|}
\hline
$i$&$p_i$&$q_i$\\
\hline
2	& $9.70703681$	&$14.5187212$\\
4	& $3.7357\times10^{-4}$	&$40.88333$\\
6	& $-1.43573$&$2.90966$\\
8	& $6.0\times10^{-8}$	&$99.999998$\\
10	& $ 1.19052066$	&$-1.11542229 $\\
12	& $9.9527277$	&$4.579 \times10^{-5}$\\
14	& $2.5455841\times10^{-1}$	&$3.866171  \times10^{-2}$\\
16	& $12.7977739$	&$10.3580447$\\ \hline
\end{tabular}
\end{center}
\end{table}

We also need a reliable estimate of the experimental uncertainties in the form
factors, in order to determine the uncertainty in the extracted coefficients
$\widetilde{c}_n$.  In the global analysis, there are two sources that can
contribute to the uncertainties in $G_E$ and $G_M$: the uncertainty on 
each individual cross section or polarization ratio, and the normalization
uncertainty associated with each cross section data set.  The normalization
factors are allowed to vary in the fit, as was the case in
Ref.~\cite{Arrington:2007ux}.  To estimate the uncertainty in the fitted
normalization factors, we take the normalization factor from a single data set
and vary it around its best fit value (while allowing all other parameters to vary) 
 to map out the change in the $\chi^2$ of
the fit as a function of the normalization factor.  This yields uncertainties
between 0.2\% and 2.5\% (typically 0.6\%--1\%), compared to the initally
quoted uncertainties of 1.5\% to 5\%, for the data before the normalization has
been constrained by the fit.  However, by assuming that all 
uncertainties are entirely uncorrelated or pure normalization factors, we
neglect the possibility there may be some angle-dependent or $Q^2$-dependent 
correction that could bias the determination of the relative normalization
coefficients.  Thus, we assume that the final uncertainty on each normalization
factor is at least 0.5\%, even if the result of the $\chi^2$ analysis yields
a smaller result.

Having the uncorrelated uncertainties for each data point and the constrained
normalization uncertainties, we then extract the uncertainties for $G_E$ and
$G_M$.  For the uncorrelated uncertainties, we randomly shift each cross
section and polarization ratio measurement within its uncertainties, and
then redo the fit for $G_E$ and $G_M$.  We repeat this 1000 times,
and look at the range of values for several $Q^2$ values (55 $Q^2$ values
between 0.007 and 31.2~GeV$^2$).  This yields our uncorrelated uncertainty
at each of the $Q^2$ points.  To obtain the impact of the normalization
uncertainties, we repeat this procedure, varying the normalization of each
cross section data set according to its uncertainty, and determine the range
of $G_E, G_M$ values for the same set of $Q^2$ points.  In this procedure, the
uncertainty obtained depends on the fit function used, and a functional form
with insufficient flexibility will yield significant smoothing of the results
and thus unrealistically small uncertainties.  We scale up our uncertainties by
a factor of two, which yields good agreement with best direct measurements of
the form factors and uncertainties. 

As mentioned above, we use the electric and magnetic radii extracted from just
the low $Q^2$ data in as a constraint to the global fit, which can yield 
unrealisically small uncertainties for below $Q^2$=0.2~GeV$^2$, especially
for $G_M$, where the very low $Q^2$ data is extremely limited.  Thus, for these
low $Q^2$ values, we calculate the uncertainty at each $Q^2$ corresponding
to the uncertainty in the extracted radius, assuming the linear expansion. We
take this larger uncertainty, rather than the result from the fit, until the
uncertainties from direct extractions of the form factors are of comparable
size, at which point we take the direct extraction of the uncertainty. 
For $Q^2>10$~GeV$^2$, there are no direct extractions of $G_E$, and
thus we again have to be sure that we do not underestimate the uncertainties.
The global fit yields $G_E/G_{D} \approx 0$ at high $Q^2$, but it is difficult
to tell if $G_E$ becomes zero, or if $G_E/G_M$ continues its linear decrease
with $Q^2$~\cite{Puckett:2010ac}.  Thus, for $Q^2>10$ GeV$^2$, we set the
uncertainty to be the difference between the best fit, which yields $G_E \approx
0$ and the fit where the linear falloff in $G_E/G_M$ continues, with $G_E$
changing sign  and  then increasing in absolute
value.

\begin{figure}[htb]
\epsfig{file=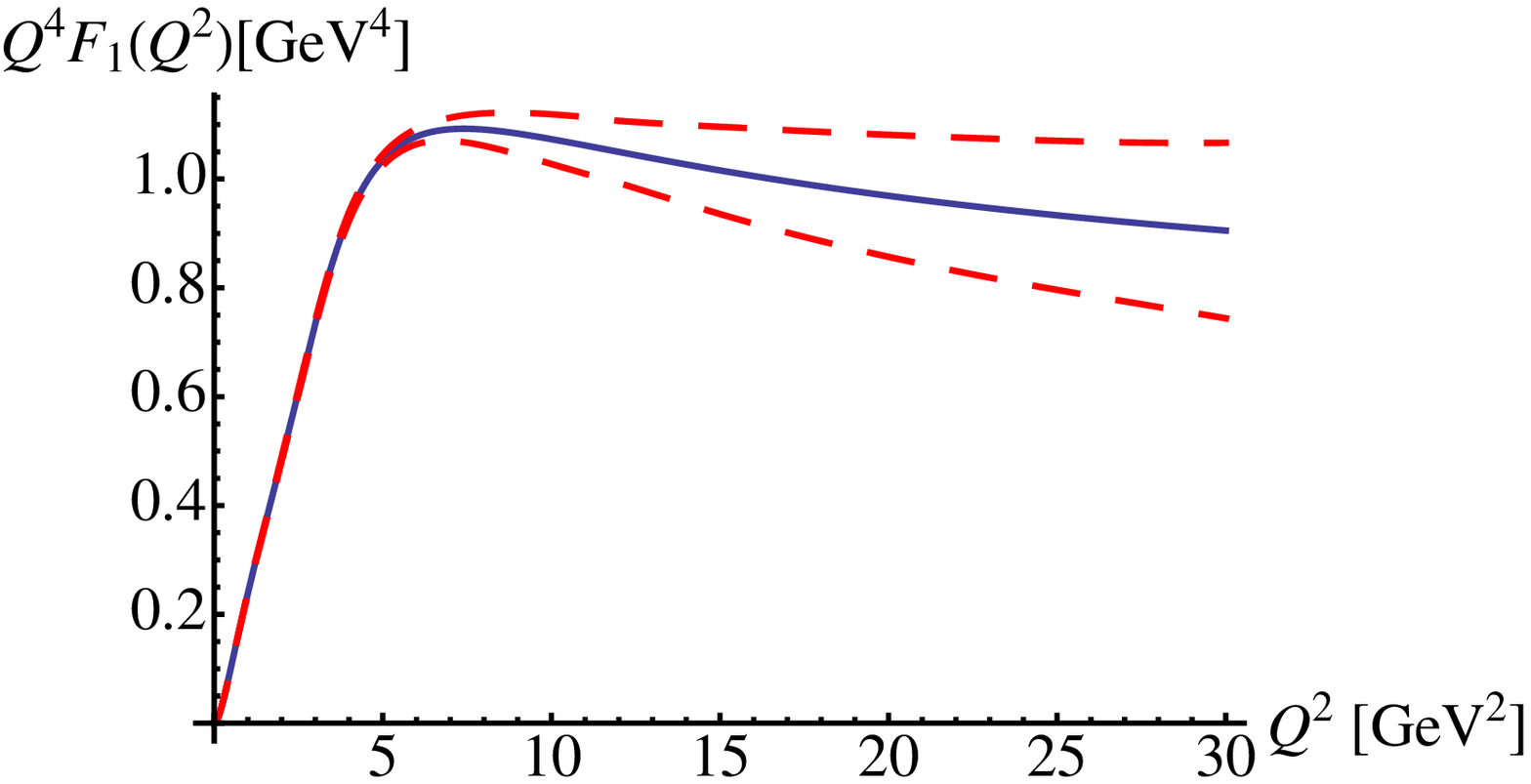, width=.48\textwidth}
\epsfig{file=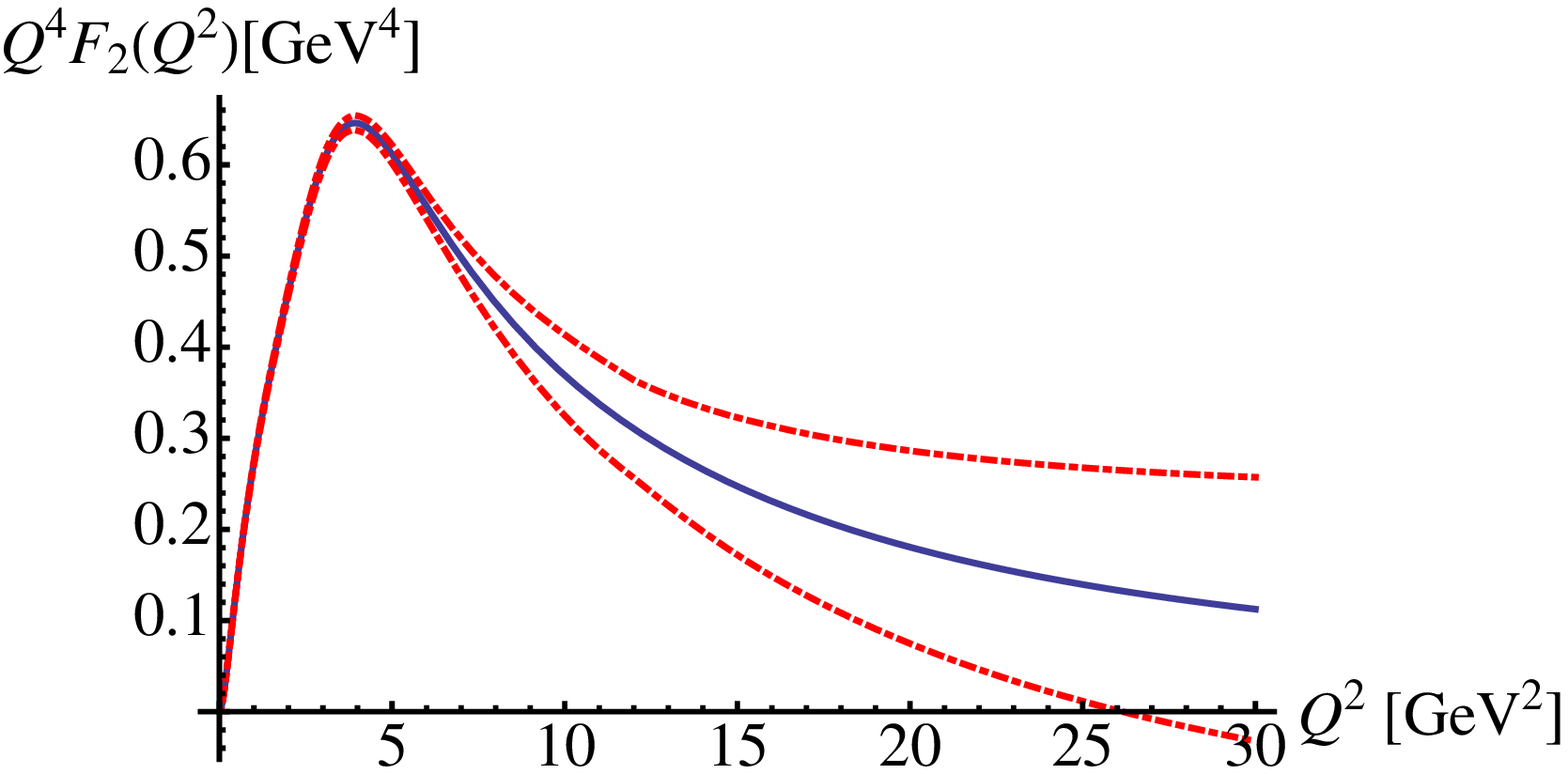, width=.48\textwidth}
\caption{(Color online) The electromagnetic form factors $F_1(Q^2)$  and $F_2(Q^2)$ and their error bands,
 scaled by a factor of $Q^4$. }
\label{F1F2Graph}\end{figure}

We then use the fit and uncertainties for $G_E$ and $G_M$ to extract
$F_1$ and $F_2$, treating the uncertainties in $G_E$ and $G_M$ as uncorrelated,
yielding:
\be
(dF_1)^2=(\frac{1}{1+\tau})^2 (dG_E)^2+(\frac{\tau}{1+\tau})^2 (dG_M)^2 \label{dF1}
\ee
\be
(dF_2)^2=(\frac{1}{1+\tau})^2 (dG_E)^2+(\frac{1}{1+\tau})^2 (dG_M)^2 \label{dF2} ~, 
\ee
While the Rosenbluth extractions yield a strong anti-correlation between the
uncertainties on $G_E$ and $G_M$, the polarization ratio yields a correlated
uncertainty; in the global fit, the combined result is fairly well approximated
by entirely uncorrelated uncertainties.
Figure~\ref{F1F2Graph} shows the extracted values of $F_1$ and $F_2$ along with
their uncertainties. Because the elastic cross section is dominated by the
contribution from $G_M$ at large $Q^2$, the fractional uncertainties on $G_E$
are much larger, and the uncertainty on $G_E$ dominates the uncertainty on
both $F_1$ and $F_2$, even though its contribution to $F_1$ is supressed by a
factor of $\tau$ relative to the $G_M$ contribution.

We note that for $Q^2<0.5$, the uncertainty coming from cross section
normalizations can be the larger contribution to the total uncertainty (and
it's dominant for $G_E$ below 0.1 GeV$^2$). While the normalization
uncertainty in the cross sections won't give a normalization style uncertainty
on $G_E$, the normalization of a given experiment will tend to have a
correlated effect on all of the extractions within the $Q^2$ covered by the
experiment. This effect is accounted for by using the procedure discussed below in Sect.~\ref{subexp}.

\section{Extraction of Realistic Proton Transverse Densities}\label{results}
The principle aim of this paper is to use data observed in experiments to obtain the charge and magnetization densities.
Recall that the transverse charge density $\rho_{ch}$ is given by 
\be
\rho_{ch}(b)=\frac{1}{2\pi }\int Q dQJ_0(Qb)F_1(Q^2). \label{rhoe}
\ee
The two-dimensional Fourier transform  of $F_2$, $\rho_{2}$ is similarly given by
\be
\rho_2(b)=\frac{1}{2\pi}\int Q dQJ_0(Qb)F_2(Q^2). \label{rho2}
\ee
However the true magnetization density, obtained by computing the expectation value of the transverse position operator with the electromagnetic current operator is given \cite{Miller:2010nz} by
\bea
\rho_m(b)&=&-b \frac{d}{db} \rho_2(b)\cr
&=&\frac{b}{2\pi}\int Q^2 dQJ_1(Qb)F_2(Q^2). \label{rhom}
\eea
This quantity is the density related to the anomalous magnetic moment. 
We begin by  extracting $\rho_{ch,2}$. The starting point is to use  the above expressions  along with the experimentally
determined $F_{1,2}$ obtained from the fits of Sect.~\ref{fitting}.
But extracting realistic transverse densities requires that a determination of
the uncertainties in the results. There are two sources of uncertainty.
Experimental data have uncertainties in the region where they are measured,
and no direct information is available above some maximum value of
$Q^2=Q^2_{max}$, where there are no measurements.  The experimental
uncertainties lead directly to uncertainties in the $\widetilde{c}_n$ via \eq{cfra}, and
can be taken into account without further ado.  However, uncertainty must
arise because of lack of knowledge of form factors for $Q^2>Q^2_{max}$, and
these need to be estimated.  This error is called the incompleteness error.

\subsection{Impact of Experimental Uncertainties on the Extracted Transverse Densities}\label{subexp}

We first treat the experimental uncertainties.  We only use the series \eq{efra} for values of $Q^2_n$  for which form factors have been extracted.  The magnetic form factor $G_M$ is well measured up to
$Q^2=31$~GeV$^2$, but $G_E$ is only known up to $\sim$10~GeV$^2$.  Based on the
estimated uncertainties on $G_E$ above 10~GeV$^2$, we find that while $F_1$ is
relatively well measured up to 30~GeV$^2$, the uncertainties on $F_2$ grow
rapidly above 10~GeV$^2$, reaching 25\% by 13~GeV$^2$.  
These upper limits on $Q^2$  are related to limits on the summation index $n$ (of \eq{efra}) through \eq{qn} which
 requires values of $R_i$.
Taking $\langle b^2 \rangle$ given by  $\rho_{ch,2}$ from the
fits presented above, we use \eq{msr} to obtain  $R_1=3.29$ fm and $R_2=3.62$ fm
for $F_{1,2}$. This corresponds to  upper limits $N$ on the sum over $n$
$n=30,\;Q_{30}^2=31\;{\rm GeV}^2$ for $\rho_{ch}(b)$, but only up to $n=20,\;Q_{20}^2=11$ GeV$^2$
 for $\rho_{2}(b)$.

\begin{figure}[htb]
\epsfig{file=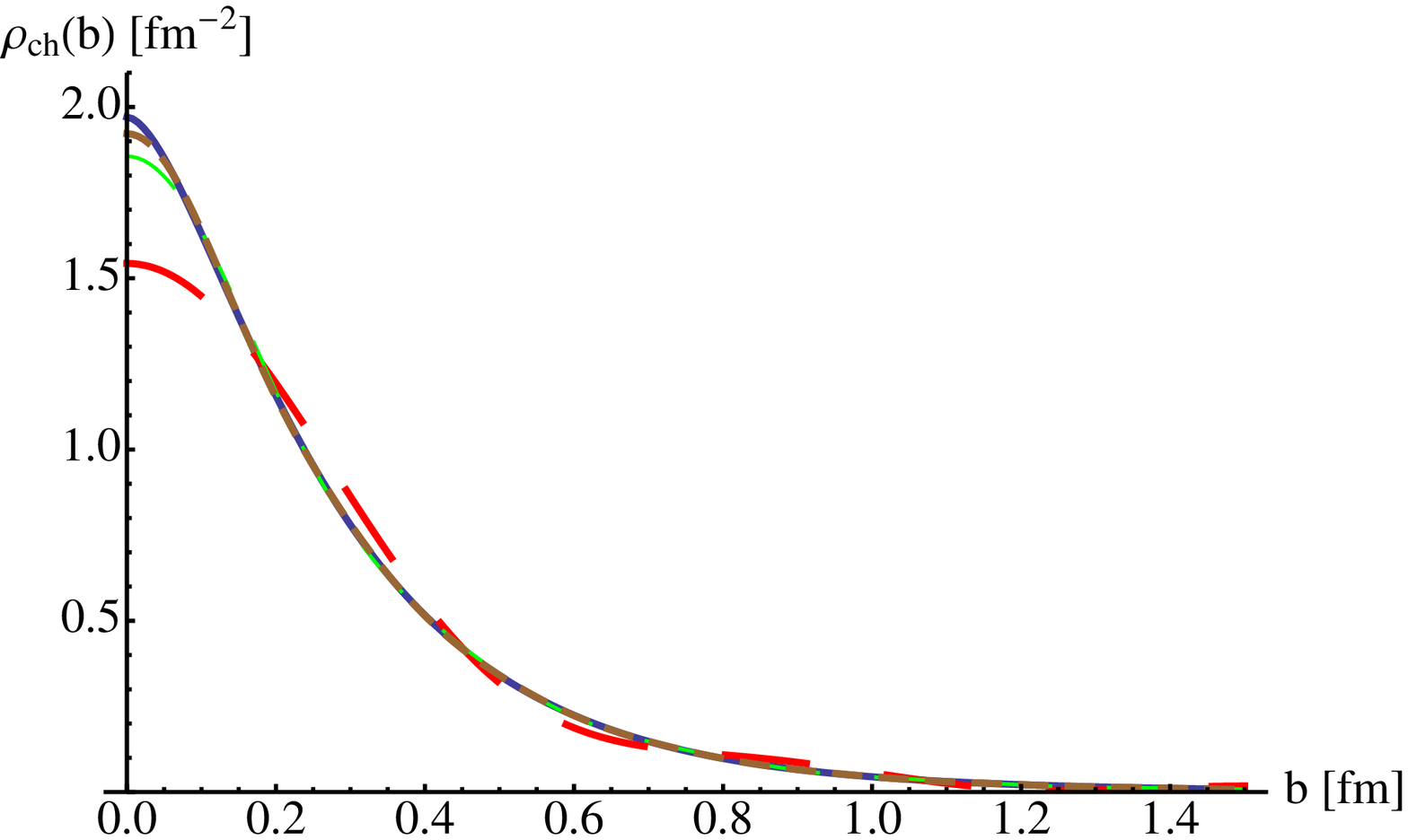, width=.48\textwidth}
\epsfig{file=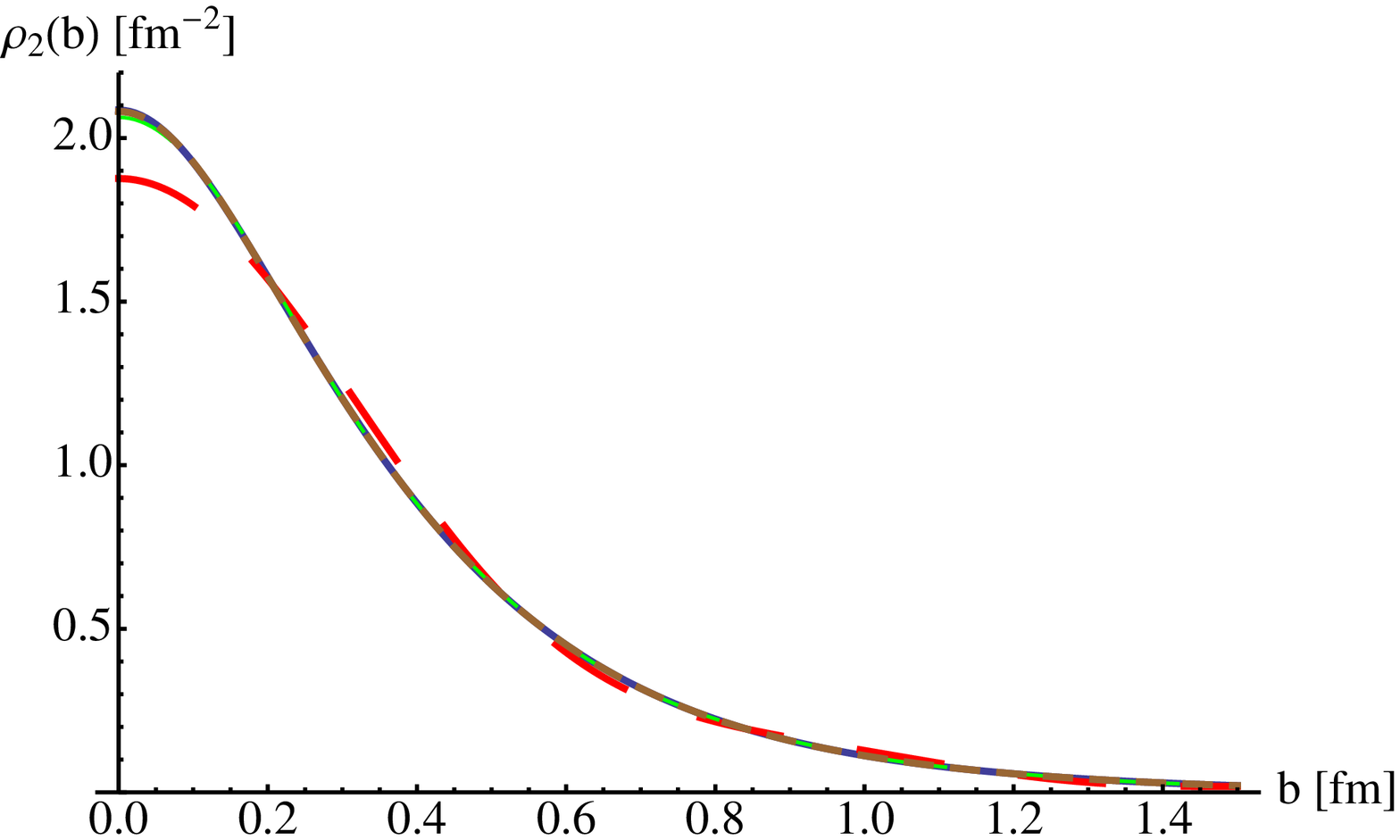, width=.48\textwidth}
\caption{(Color Online) The transverse densities $\rho_{ch}$,$\rho_2$ (blue, solid) of the parameterizations and their  approximates to  10 terms
(red,long dash), 20 terms (green, medium dash), 30 terms (brown, short dash)
and using the parameterization of \eq{GeGm}. The
approximations converge as the number of terms increases. }
\label{rhographs}
\end{figure}

\begin{figure}[htb]
\epsfig{file=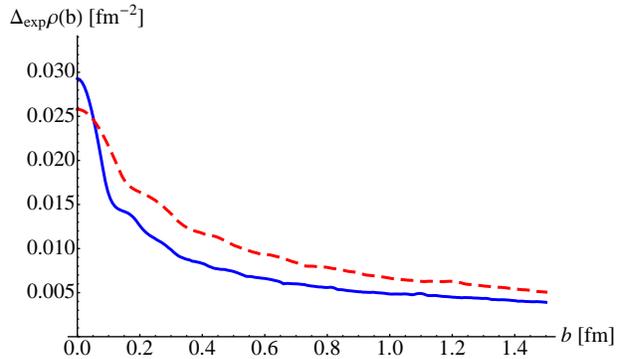, width=.48\textwidth}
\caption{(Color online) Uncertainties in transverse densities $\Delta_{exp}\rho_{ch}$ (solid, blue)
 and $\Delta_{exp}\rho_2(b)$ (dashed, red)
due to experimental uncertainties on $F_1$, $F_2$.}
\label{deltarhos}\end{figure}

The transverse densities $\rho_{ch,2}$ are plotted as the solid curves in Fig.~\ref{rhographs}. The densities peak at $b=0$ and that the transverse density $\rho_2$ has a slightly broader spatial extent than that of $\rho_{ch}$.

The next step is to 
extract $\widetilde{c}_n$ from the fit to the form factor using \eq{cfra}.
The uncertainty on $F_{1,2}(Q^2)$ directly yields an uncertainty on
$\widetilde{c}_n$, and thus its contribution to $\rho(b)$ (\eq{exp}).
Assuming the errors from each $\widetilde{c}_n$ extraction add constructively,
we obtain
\bea&&
\Delta_{exp}\rho_{ch}(b)=\sum_{n=1}^{30}\left|\frac{\partial\rho_{ch}(b)}{\partial F_1}\right|dF_1((\frac{X_n}{R_1})^2) \nn
  &&     =\frac{1}{\pi R_1^2}\sum_{n=1}^{30}J_1(X_n)^{-2}\left|J_0(X_n\frac{b}{R_1})\right|dF_1((\frac{X_n}{R_1})^2),\nn&&\label{dpe}
\eea
\bea&&
\Delta_{exp}\rho_2(b)=\sum_{n=1}^{20}\left|\frac{\partial\rho_2(b)}{\partial F_2}\right|dF_2((\frac{X_n}{R_2})^2) \nn
   &&    =\frac{1}{\pi R_2^2}\sum_{n=1}^{20}J_1(X_n)^{-2}\left|J_0(X_n\frac{b}{R_2})\right|dF_2((\frac{X_n}{R_2})^2).\nn&&\label{dpm}
\eea
Note that the errors are added linearly. This means that we are taking the worst case possible by assuming a full correlation.
These uncertainties in densities are plotted in Fig.~\ref{deltarhos}. They are
about 1.5\% of the transverse density at $b=0$ and decrease (in absolute value)
at increasing distances.  The fractional uncertainty is small (below 10\%)
until $b \approx 1$~fm, where the density is only a few percent of the peak
density.

%

\subsection{Incompleteness Error}

We next study  the uncertainties in the transverse density caused by lack of experimental knowledge at large values of $Q^2$. The first step is to understand the meaning of the truncations made in \eq{rhoe} and \eq{rhom}.
Plots of these approximations are given in \fig{rhographs}. We see that for,
$\rho_2$, one achieves agreement with the parameterization for values of $N$
as low as 20, with the largest disagreement at $b=0$. For $\rho_{ch}(b=0)$, the
difference between the result from the parameterization and the $N=30$
approximation is -2\%, while for $\rho_2(b=0)$, the $N=20$ approximation is
only 1\% below the full result.  Even though fewer terms are included in the
approximation for $F_2$, the agreement is comparable, due to the more rapid
fall-off of $F_2$ with increasing values of $Q^2$.

\begin{figure}[htb]
\epsfig{file=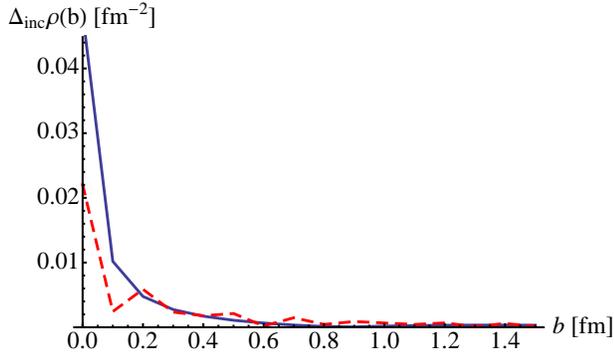, width=.48\textwidth}
\caption{(Color online) Incompleteness error.
The absolute error in $\rho_{ch}$ (solid, blue) and $\rho_{2} $ (dashed, red).}
\label{incgraphs}\end{figure}

Given this information, we can state our procedure. Our basic transverse densities are obtained by using the parameterization \eq{GeGm} to evaluate the expressions of \eq{rhoe},\eq{rho2}, and \eq{rhom}.
However, we are justified in using this parameterization for values of $Q^2$ corresponding to $N=30,(20)$ for $F_{1,(2)}$.  We assume a maximum error by taking  the uncertainty  in the form factor to be $\pm$ the value given by the parameterization. Therefore 
the estimated incompleteness uncertainty is given 
 by the expression
\bea
\Delta_{inc}(b)\equiv \left| \sum_{N+1}^\infty c_nJ_0(X_n/R_1)F_i(Q_n^2) \right| ~,\label{inc}
\eea
as a function of $b$, with $i=1,2$.  The results are shown in Fig.~\ref{incgraphs}. It is necessary to realize that using this
expression for the incompleteness error overestimates the error because
using this expression is equivalent to assuming that the form factor vanishes
for $Q^2>Q^2_N$ in \eq{efra}.  But  the form factor can  not suddenly drop to 
0.
Fig.~\ref{F1F2Graph} shows  a fractional error bar for $F_1(31 \;{\rm
GeV}^2)$ which is only about $0.2$, a fractional error bar  at 13 GeV$^2$
which is only about 0.3 of the form factor $F_2$. Thus  using \eq{inc} amounts to making an
  overestimate. To be conservative, we  obtain  
  the total uncertainty by adding the contributions of 
\eq{dpe} (or \eq{dpm}) to the estimated incompleteness uncertainty given by
\eq{inc}.

\begin{figure}[htb]
\epsfig{file=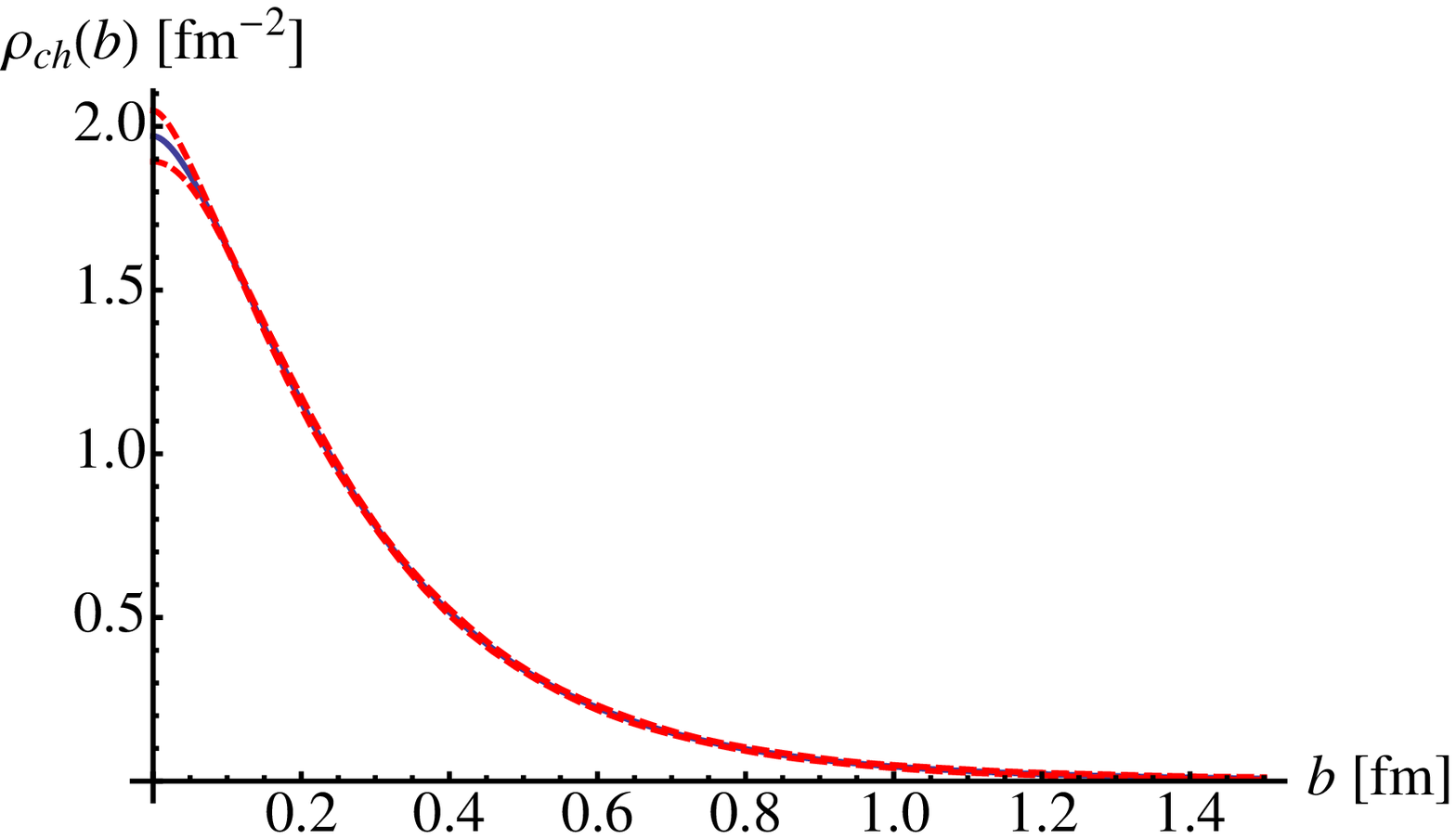, width=.48\textwidth}
\epsfig{file=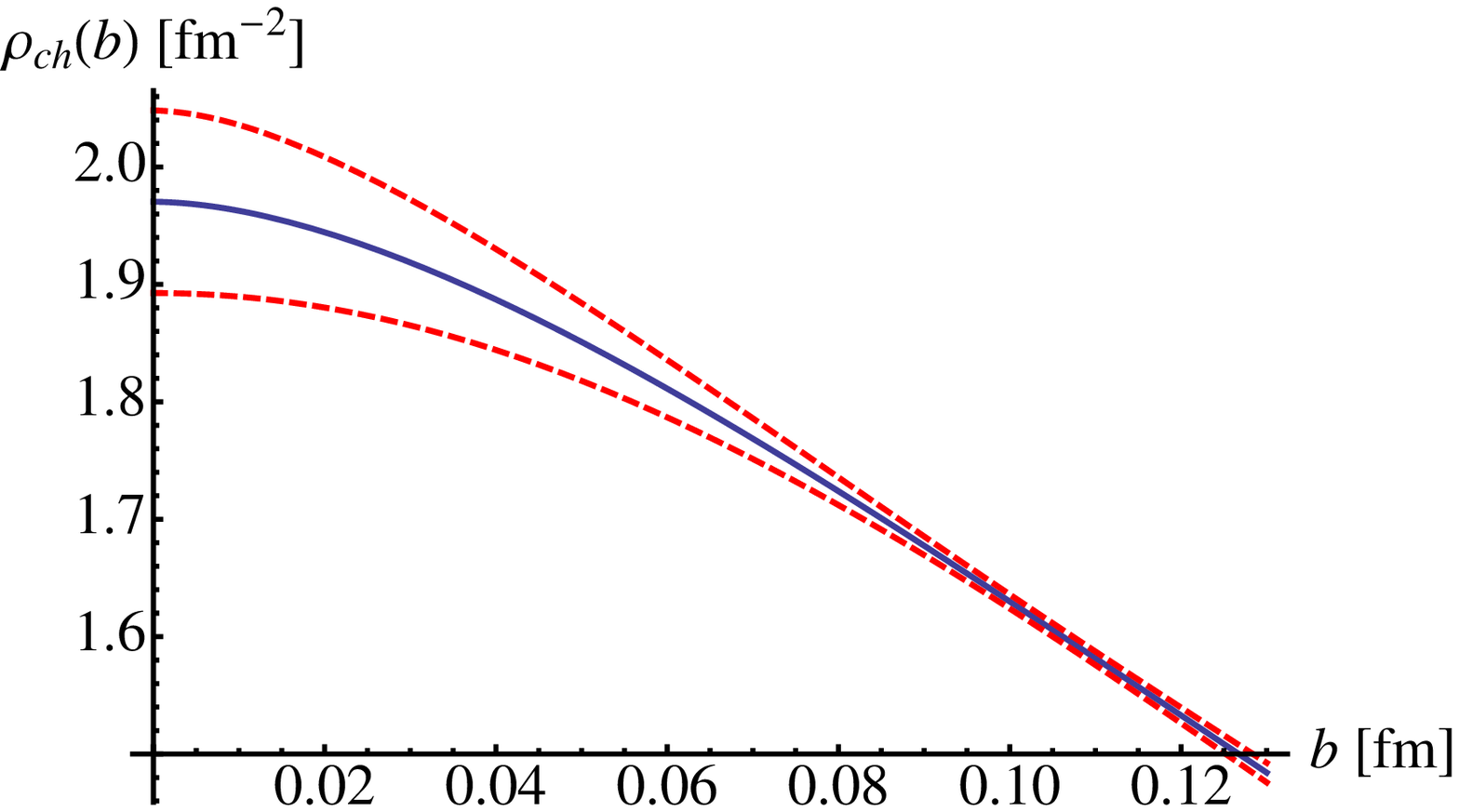, width=.48\textwidth}
\epsfig{file=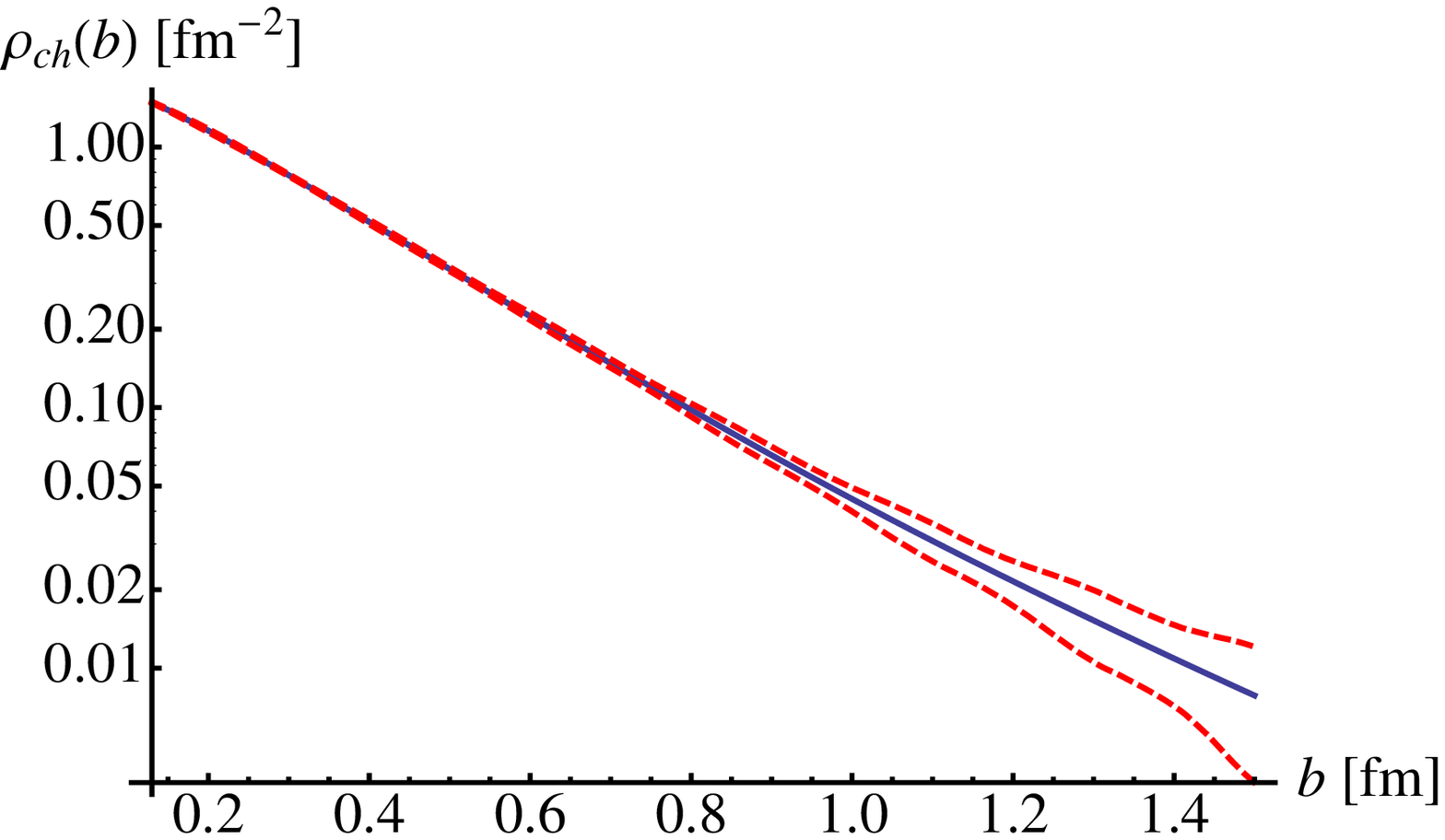, width=.48\textwidth}
\caption{(Color online) $\rho_{ch}$ (solid, blue) (with error bands (short dashed, red)).   }
\label{rhobandsch}\end{figure}

\begin{figure}[htb]
\epsfig{file=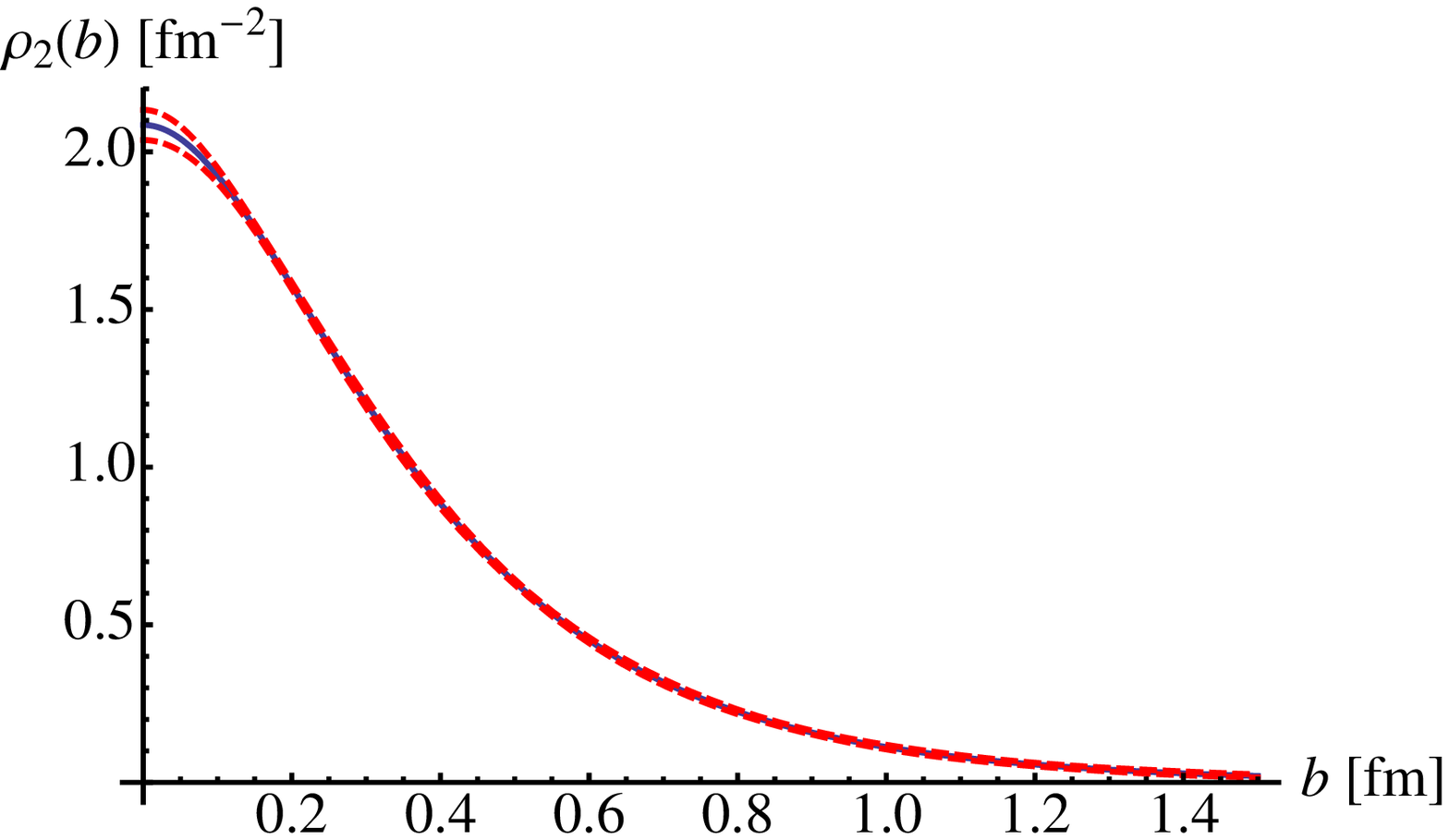, width=.48\textwidth}
\epsfig{file=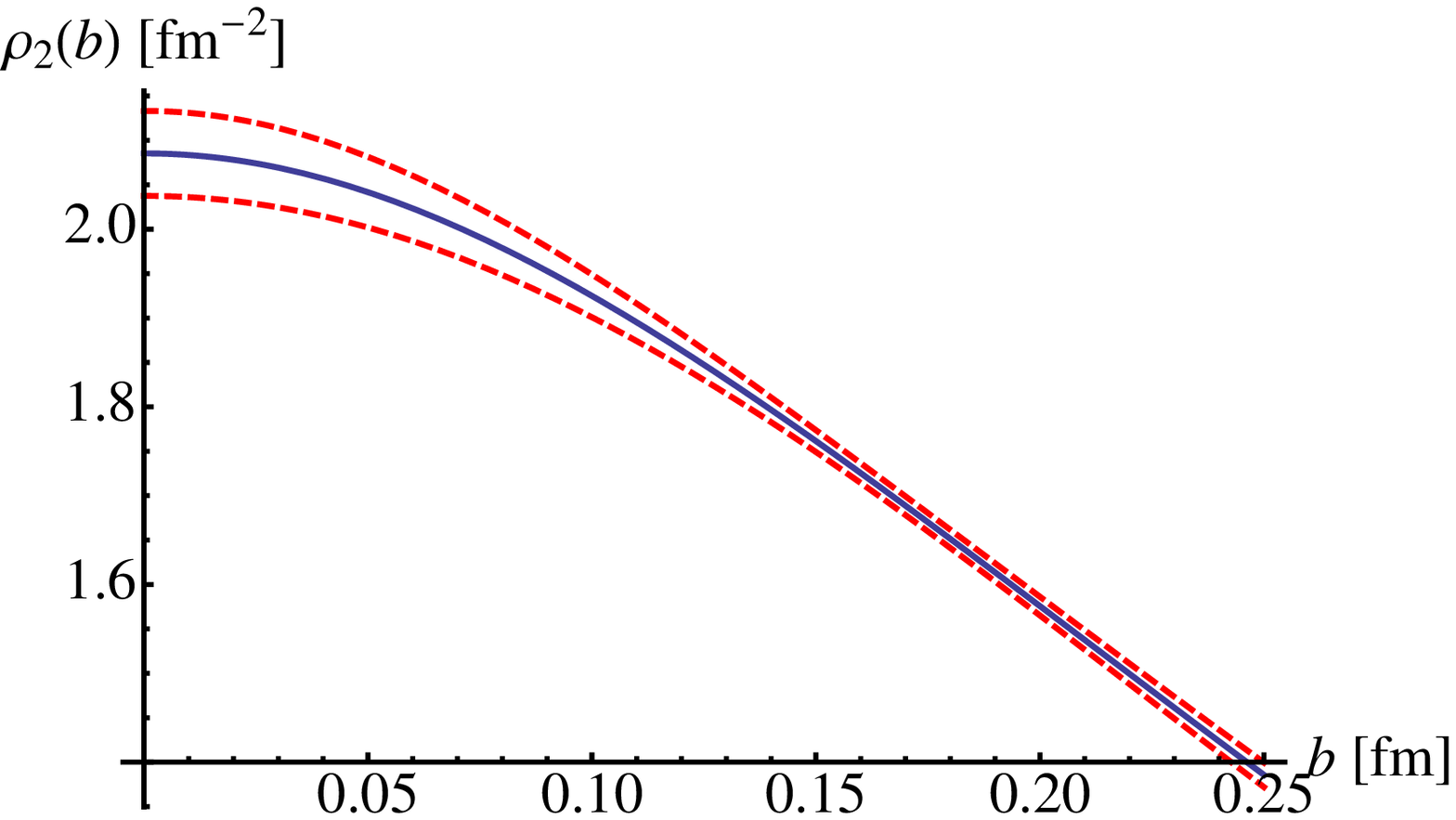, width=.48\textwidth}
\epsfig{file=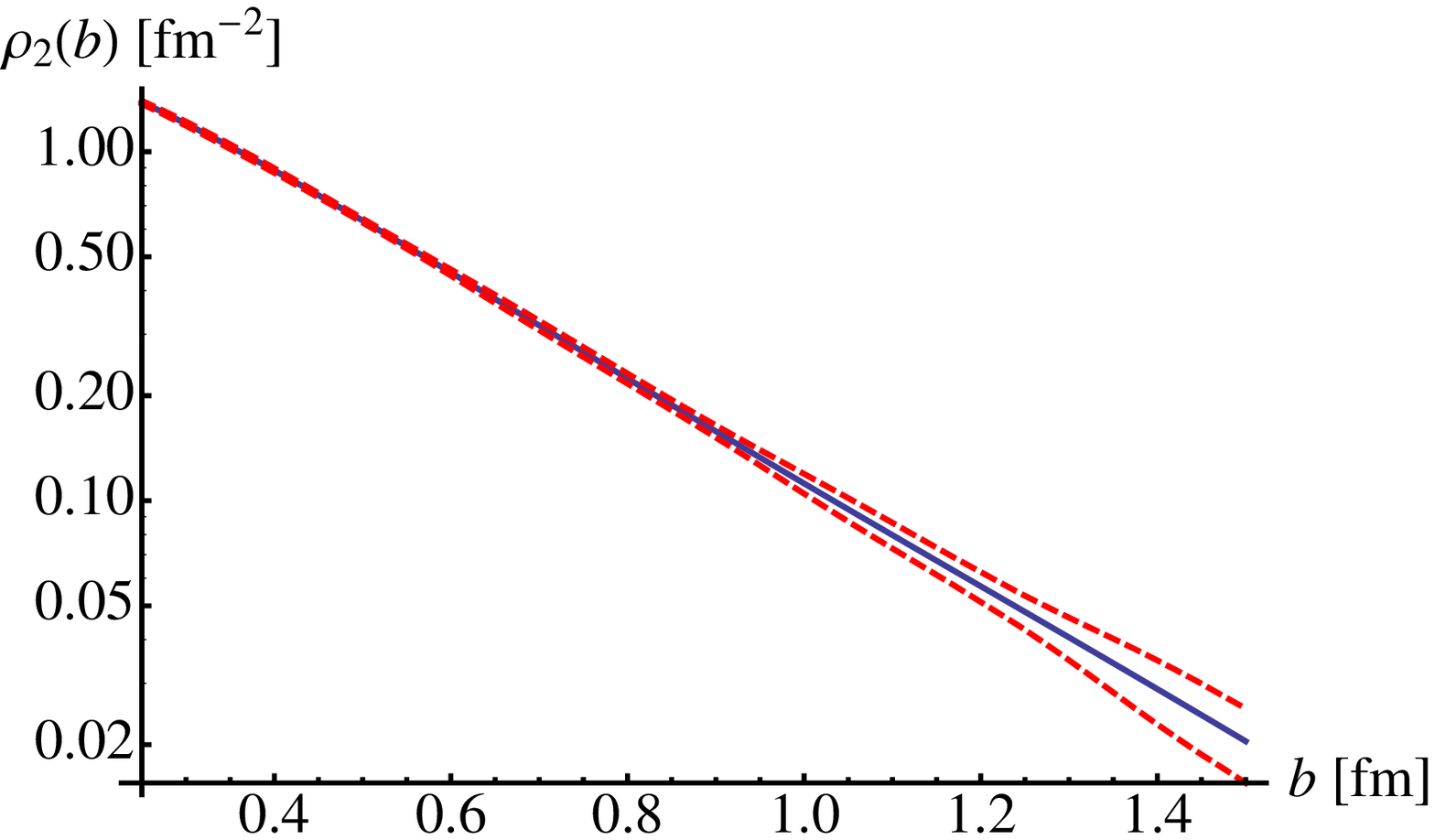, width=.48\textwidth}
\caption{(Color online) $\rho_2$, with error bands }  
\label{rhobands2}\end{figure}

We now have working expressions for the transverse densities $\rho_{ch,2}$,
and their respective uncertainties. We start with the basic  term for $\rho_{ch,2}$,
obtained by using the parameterization \eq{GeGm} to evaluate the expressions of \eq{rhoe},\eq{rho2}, and \eq{rhom},
 then add the two separate errors $\Delta_{inc,exp}$ to get a
total error $\Delta=\Delta_{inc}+\Delta_{exp}$ for $\rho_{ch}$. A band is
formed by considering the region between the basic plus or minus the
appropriate $\Delta$ for the two densities. 

The transverse densities $\rho_{ch,2}(b)$ are plotted with their error bands
in \fig{rhobandsch} and \fig{rhobands2}. The errors are very small except for
values of $b$ less than about 0.1~fm.  The results in this figure are the
central numerical findings of this paper. The transverse densities are known
very well indeed. The spatial extent of $\rho_2$ is broader than that of
$\rho_{ch}$ as previously observed \cite{Miller:2007kt}.
Note that the realistic transverse densities differ substantially from the 
 dipole result of \eq{rhoD}, shown in  Fig.~2. 

\begin{figure}[htb]
\epsfig{file=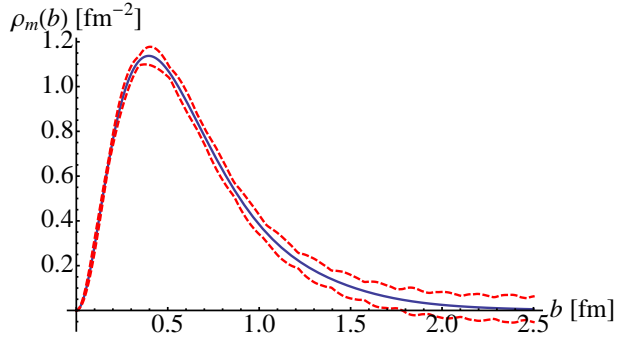, width=0.48\textwidth}
\caption{(Color online) The true magnetization density $\rho_m$. The
uncertainties are numerically negligible. }
\label{drhographs}\end{figure}

\subsection{Extraction of $\rho_m(b)$}\label{subsrhom}
We now turn to the true transverse anomalous magnetic density of \eq{rhom}, defined by taking the matrix element of 
${1\over 2}\int\;d^3 r \bfb \times \vec{j}$ in a transversely polarized state,\cite{Miller:2010nz,Miller:2007kt}. This Fourier transform involves $J_1(Qb)$
and therefore the FRA corresponds to that of \eq{gen} and \eq{gencn}, with $\lambda=1$.  Using this expansion, instead of simply taking the derivative of $\rho_2$, allows an expansion in basis functions that explicitly vanish at $b=R_2$. 
Then the FRA  gives the result:
\bea&&
\rho_m=
\frac{1}{\pi R_2^2}\sum_{n=1}^{\infty}J_2^{-2}(X_{1,n})b Q_{1,n}F_2(Q_{1,n}^2)J_1(Q_{1,n}{b}),\cr 
&&Q_{1,n}\equiv {X_{1,n}\over R_2}.
\label{rhomtrue}
\eea
Once again we include the effects of the experimental error and the
incompleteness error. This latter error is larger in this case than for
$\rho_2$ because of the explicit factor of  $X_{1,n}$. The result for $\rho_m$
and its error bands are  plotted in \fig{drhographs}.  This quantity has a
broader spatial extent than $\rho_2$, possibly resulting from the importance
of the pion cloud in causing the anomalous magnetic moment. The uncertainties
on this quantity are greater than for the other densities. Future measurements
extending knowledge of $F_2$ to higher values of $Q^2$ would reduce these
higher uncertainties.

\section{Summary}\label{summary}

This paper is concerned with obtaining a general method to determine information about densities in the transverse plane. The use of Bessel series expansion, augmented by the finite radius approximation FRA of \eq{exp},
\eq{cfra}, \eq{gen} and \eq{gencn}
   allows us to determine the effects of experimental uncertainties and also allows us to estimate the effects of the incompleteness error caused by  a lack of measurements  at large values of $Q^2$. 
  The method can be applied to the extraction of any spatial quantity. One example, related to orbital angular momentum,  is shown    in  \eq{gen} and \eq{gencn}.

   The method is applied here to analyze electromagnetic form factors. 
We can see from \fig{rhobandsch} and  \fig{rhobands2} that the errors associated with the transverse charge density and the two-dimensional Fourier transform of $F_2$ are 
 very small.  The anomalous magnetization density $\rho_M$, \fig{drhographs},  is also reasonably well determined, but future  measurements extending our knowledge of $F_2$ to higher values of $Q^2$    would reduce the existing uncertainties. 
 \section*{Acknowledgments}
This research was supported by the NSF REU program, grant PHY-0754333, and the USDOE grants
FG02-97ER41014
and  DE-AC02-06CH11357.  GAM wishes to thank Jefferson Laboratory for its hospitality during a visit while this work was being completed. We thank A.~Puckett and M.~Diehl
for useful comments on the manuscript.


\begin{thebibliography}{99}
\bibitem{Gao:2003ag}
  H.~y.~Gao,
  Int.\ J.\ Mod.\ Phys.\  E {\bf 12}, 1 (2003)
  [Erratum-ibid.\  E {\bf 12}, 567 (2003)]
\bibitem{HydeWright:2004gh}
  C.~E.~Hyde and K.~de Jager,
  Ann.\ Rev.\ Nucl.\ Part.\ Sci.\  {\bf 54}, 217 (2004)
\bibitem{Perdrisat:2006hj}
  C.~F.~Perdrisat, V.~Punjabi and M.~Vanderhaeghen,
  Prog.\ Part.\ Nucl.\ Phys.\  {\bf 59}, 694 (2007)
\bibitem{Arrington:2006zm}
  J.~Arrington, C.~D.~Roberts and J.~M.~Zanotti,
  J.\ Phys.\ G {\bf 34}, S23 (2007)


\bibitem{Horn:2007ug}
  T.~Horn {\it et al.},
  Phys.\ Rev.\  C {\bf 78}, 058201 (2008)
\bibitem{Blok:2008jy}
  H.~P.~Blok {\it et al.}  [Jefferson Lab Collaboration],
  Phys.\ Rev.\  C {\bf 78}, 045202 (2008)



 \bibitem{Hofstadter:1956qs}
  R.~Hofstadter,
  Rev.\ Mod.\ Phys.\  {\bf 28}, 214 (1956).
\bibitem{Miller:2009sg}
  G.~A.~Miller,
  Phys.\ Rev.\  C {\bf 80}, 045210 (2009)


\bibitem{Miller:2007uy}
  G.~A.~Miller,
  Phys.\ Rev.\ Lett.\  {\bf 99}, 112001 (2007)

\bibitem{Miller:2010nz}
  G.~A.~Miller,
  arXiv:1002.0355 [nucl-th], to be published ARNPS, 2010.

\bibitem{Burkert:2008rj}
  V.~D.~Burkert,
  arXiv:0810.4718 [hep-ph].

\bibitem{SN} H. Nyquist,  Trans. 
Amer. Inst. Elect. Eng.,  {\bf 47},  617 (1928); C. E. Shannon,  Proc. 
IRE, {\bf 37}, 10 (1949).  

\bibitem{Soper:1976jc}
  D.~E.~Soper,
  Phys.\ Rev.\  D {\bf 15}, 1141 (1977).


\bibitem{Strikman:2010pu}
  M.~Strikman and C.~Weiss,
  arXiv:1004.3535 [hep-ph].

\bibitem{Jackson}
  J.~D.~Jackson, {\it Classical Electrodynamics Third Edition} (Wiley, New York, 1998)  
\bibitem{Blunden:2005ew}
  P.~G.~Blunden, W.~Melnitchouk and J.~A.~Tjon,
  Phys.\ Rev.\  C {\bf 72}, 034612 (2005)
  [arXiv:nucl-th/0506039].

\bibitem{Arrington:2007ux}
  J.~Arrington, W.~Melnitchouk and J.~A.~Tjon,
  Phys.\ Rev.\  C {\bf 76}, 035205 (2007)
  [arXiv:0707.1861 [nucl-ex]].

\bibitem{Puckett:2010ac}
  A.~J.~R.~Puckett {\it et al.},
  Phys.\ Rev.\ Lett.\  {\bf 104}, 242301 (2010)
  [arXiv:1005.3419 [nucl-ex]].

\bibitem{Miller:2007kt}
  G.~A.~Miller, E.~Piasetzky and G.~Ron,
  Phys.\ Rev.\ Lett.\  {\bf 101}, 082002 (2008)



\end{thebibliography}
\end{document}